\documentclass[pre,aps, preprint]{revtex4-1}
\usepackage{amsmath,natbib,amsfonts}
\usepackage{amssymb,color}
\usepackage{graphicx}
\usepackage{subcaption,enumerate}
\newcommand{\nwc}{\newcommand}
\nwc{\vs}{\vspace}
\nwc{\hs}{\hspace}
\nwc{\la}{\langle}
\nwc{\ra}{\rangle}
\nwc{\nn}{\nonumber}
\nwc{\Ra}{\Rightarrow}
\nwc{\wt}{\widetilde}
\nwc{\lw}{\linewidth}
\nwc{\ft}{\frametitle}
\nwc{\ben}{\begin{enumerate}}
\nwc{\een}{\end{enumerate}}
\nwc{\bit}{\begin{itemize}}
\nwc{\eit}{\end{itemize}}
\nwc{\dg}{\dagger}
\nwc{\mA}{\mathcal A}
\nwc{\mD}{\mathcal D}
\nwc{\mB}{\mathcal B}
\nwc{\pd}[2]{\frac{\partial #1}{\partial #2}}
\nwc{\av}[1]{\left< #1\right>}
\nwc{\alert}[1]{\textcolor{red}{#1}}

\begin{document}

\title{Convergence of thermodynamic quantities and work fluctuation theorems in presence of random protocols}
\author{Rahul Marathe}
\email[]{maratherahul@physics.iitd.ac.in}
\affiliation{Department of Physics, Indian Institute of Technology, Delhi, Hauz Khas 110016, New Delhi, India.}
\author{Sourabh Lahiri}
\email[]{sourabhlahiri@gmail.com}
\affiliation{Department of Physics, Birla Institute of Technology Mesra, Ranchi, Jharkhand 835215, India.}

\date{\today}
\begin{abstract}
Recently many results namely the Fluctuation theorems (FT), have been discovered for systems arbitrarily away from equilibrium. Many of these relations have been experimentally tested. The system under consideration is usually driven out of equilibrium by an external time-dependent parameter which follows a particular {\it protocol}. One needs to perform several iterations of the same experiment in order to find statistically relevant results. Since the systems are microscopic, fluctuations dominate. Studying the convergence of relevant thermodynamics quantities with number of realizations is also important as it gives a rough estimate of number of iterations one needs to perform. In each iteration the protocol follows a predetermined {\it identical/fixed} form. However, the protocol itself may be prone to fluctuations. In this work we are interested in looking at a simple non-equilibrium system namely a Brownian particle trapped in a harmonic potential. The center of the trap is then dragged according to a protocol. We however lift the condition of fixed protocol. In our case the protocol in each realization is different. We consider one of the parameters of the protocol as a random variable, chosen from some known distribution. We study the systems analytically as well as numerically. We specifically study the convergence of the average work and free energy difference with number of realizations. Interestingly, in several cases, randomness in the protocol does not seem to affect the convergence when compared to fixed protocol results. We study symmetry functions. A Brownian particle in a double well potential is also studied numerically. We believe that our results can be experimentally verified. 
\end{abstract}  

\maketitle

\section{Introduction}

The field of non-equilibrium statistical mechanics, especially when the system is far from equilibrium, has attracted a lot of works in the last two decades or so. It is of great practical importance, given the fact that in nature, one almost always deals with irreversible or non-equilibrium processes. The framework of equilibrium thermodynamics is not so useful to study such processes, and we need separate recipes for dealing with them. One of the major developments in this area has been a group of relations collectively called ``Fluctuation Theorems''. They constitute relations that have the rare distinction of being exact equalities, no matter how far the system has been driven away from equilibrium. Let the system be initially at equilibrium with a thermal bath of inverse temperature $\beta$ at a given initial value of an externally controlled parameter or \emph{protocol} $\lambda_0$. At time $t=0^+$, the time dependence of this parameter is switched on, as a result work is done on the system. Such a process is called \emph{forward process}, in contrast to the \emph{reverse process}, in which the initial value of parameter is $\lambda(\tau)$ at which the system is at thermal equilibrium,  and thereafter the reverse protocol $\lambda(\tau-t)$ is applied for the same time of observation $\tau$. Two well-known equalities hold for such processes:
\begin{align}
  \left<e^{-\beta W}\right> &= e^{-\beta\Delta F}; \hspace{2cm}\mbox{(Jarzynski Equality)}\label{JE} \\
  \frac{P(W)}{\tilde P(-W)} &= e^{\beta(W-\Delta F)}. \hspace{2cm}\mbox{(Crooks Theorem)} \label{CFT}
\end{align}
Here, $W$ is the stochastic work done along individual trajectories, and the average (denoted by angular brackets) is the ensemble average.
The first equality was proved by C. Jarzynski \cite{jar97_prl} and the second by G. E. Crooks \cite{cro98_jsp, cro99_pre}.

A large volume of works followed thereafter. In the meantime, Sekimoto established the field of \emph{Stochastic Thermodynamics} on a firm ground, thereby providing clear definitions for thermodynamic quantities along individual phase space trajectories (i.e. in individual experiments) \cite{sek98}. Fluctuation theorems for total entropy production were derived by Seifert \cite{sei05_prl, sei08_epjb}. Fluctuation theorems for systems undergoing transitions from one non-equilibrium steady state to another were studied \cite{hat01_prl, spe07}. FTs have been derived for systems following Hamiltonian as well as stochastic dynamics \cite{sei12_rpp}. They have been proved for both closed and open quantum systems \cite{han11_rmp}.

Our focus in this article would however be on the convergence of thermodynamic quantities like work, free-energy and Jarzynski and the Crooks work theorems, Eq. (\ref{JE}) and Eq. (\ref{CFT}). Till now, all the works have studied FTs with a \emph{fixed} protocol. In other words, although we are repeating the experiment a large number of times, the functional form $\lambda(t)$ is the same for all these experiments. The stochasticity of work arises entirely because the system traces out a different trajectory in phase space every time because of thermal noise, and not because of variations in protocol. 

We, in this work, would like to lift the condition of deterministic protocol and check whether the theorems are robust even if the protocol is different in different experiments of the ensemble. The parameters used in the protocols will be sampled from some distribution.
Our objective would be to compare the effect of three such distributions: the uniform, Gaussian and exponential, and to numerically observe which option provides better convergence to analytical values. In fact, it is important to state that we use the \emph{same number of realizations} of the experiment as in the case of a fixed protocol to obtain the value of $\Delta F$.

Our protocols will mainly consist of dragging a colloidal particle through a medium, using different types (time dependence) of protocols as well as using different random distributions of the parameters involved for the same functional form of the protocol. Several experimental and theoretical works have been performed using dragged colloidal particles. Wang {\it et. al.} performed an experiment \cite{wang02_prl} with a dragged colloidal particle, where he demonstrated the validity of the Crooks work fluctuation theorem. Subsequently, the authors of \cite{zon03_pre} explicitly showed the validity of the Crooks fluctuation theorem and the violation of the steady state fluctuation theorem in such systems. The model has also been studied in detail in \cite{tre04_pnas,jar99_arxiv,spe05_epjb,lah09_pre}.

We have also used sinusoidal protocols for two examples where the work distribution becomes non-Gaussian.

\section{The model}

We will consider a particle in a medium at temperature $T$, that is placed in a harmonic trap of stiffness constant $k$ and follows the overdamped Langevin equation of motion:
\begin{eqnarray}
  \gamma \dot x = -k(x - \lambda(t)) + \eta(t).
  \label{lang_drag}
\end{eqnarray}
Here, $\eta(t)$ is the usual Gaussian white noise, having the properties $\av{\eta(t)}=0$ and $\av{\eta(t)\eta(t')} = 2\gamma k_BT\delta(t-t')$, $\gamma$ is the friction coefficient of the medium, and overhead dot implies total time derivative. The center of the harmonic trap is dragged according to a protocol $\lambda(t)$.
For the analysis carried below we fix the values
of parameters to be $k_BT=1$, $\gamma =1$, $k=1$. To arrive at these dimensionless parameters, the time has been scaled by the relaxation time $\gamma/k$, the position by the square root of initial thermal width $\sqrt{k_BT/k}$, and energies are measured in units of $k_BT$, i.e. the following replacements have been made:
\begin{align}
 t \to \frac{kt}{\gamma}, \hspace{1cm} x\to \left(\sqrt{\frac{k}{k_BT}}\right)x, \hspace{1cm} W\to \frac{W}{k_B T}. \nonumber
\end{align}
The values of these parameters in any system of units can therefore be obtained by multiplying the dimensionless values by the appropriate factors listed above. The typical dimension of position at mesoscopic scales is nanometers, of time is seconds and of energy is piconewton-nanometers.

The system is initially allowed to equilibrate at $\lambda_0=0$ (the initial value of the dragging protocol) and then the external time-dependent perturbation is switched on. Note that for equilibrium initial distributions, the ergodicity of the system prevails throughout the time of observation, so that in simulations one can use time averaging if that is computationally less expensive as compared to ensemble averaging \cite{tha18_pre}, a fact that can often come handy in simulations.
The final value of the perturbation (dragging protocol), $\lambda_\tau$, will be chosen from a random distribution, thus making the protocols random in nature.

We consider two forms of the protocols:
\begin{enumerate}
\item {\bf Ramp protocol:} $\lambda(t)$ varies linearly from $\lambda_0$ at time $t=0$ to the final value $\lambda_\tau$ at time $t=\tau$. The form is given by
\begin{eqnarray}
  \lambda(t) &= \lambda_0 + (\lambda_\tau-\lambda_0)\frac{t}{\tau}.
               \label{form1}
\end{eqnarray}
The parameter $\lambda_\tau$ is a random number, sampled from a given distribution, while
without loss of generality, we chose $\lambda_0=0$.

\item {\bf Optimal protocol:} The other form that we consider is the optimal protocol described in \cite{sei07_prl}. This protocol is important since it minimizes
the work done between the given initial and the final protocol values.
\begin{eqnarray}
  \lambda(t) &=\left(\frac{t+1}{\tau+2}\right) \lambda_\tau.
               \label{form2}
\end{eqnarray}
As is obvious, the parameter $\lambda(t)$ does not reach its final value $\lambda_\tau$ at time $t=\tau$, nor does its initial value $\lambda_0$ correspond to $\lambda(0)$. This means that the optimal protocol requires a jump in the value of external parameter at the initial and the final times. Once again, $\lambda_0=0$, while $\lambda_\tau$ is a random number chosen from a distribution.
\end{enumerate}
We have chosen the most probable value of $\lambda_\tau$ same as the final value reached in the fixed protocol case (with which we have provided comparisons). Throughout this article, we choose this value to be $\bar{\lambda}_\tau=5$.

Also, since $\Delta F = 0$ for \emph{any} form of the dragging protocol, we do not need to keep the initial and final values of the parameter fixed for all protocols. Here, $\lambda_\tau$ is a constant \emph{for a particular experiment}, but varies randomly from one experiment to another. The random values, as mentioned earlier, will be sampled from uniform, Gaussian and exponential distribution and the resulting outcomes would be compared. The initial value $\lambda_0$ of the parameter is always taken to be zero. 

\begin{figure*}[!htbp]
  \begin{subfigure}{0.4\linewidth}
    \centering
    \includegraphics[width=7cm]{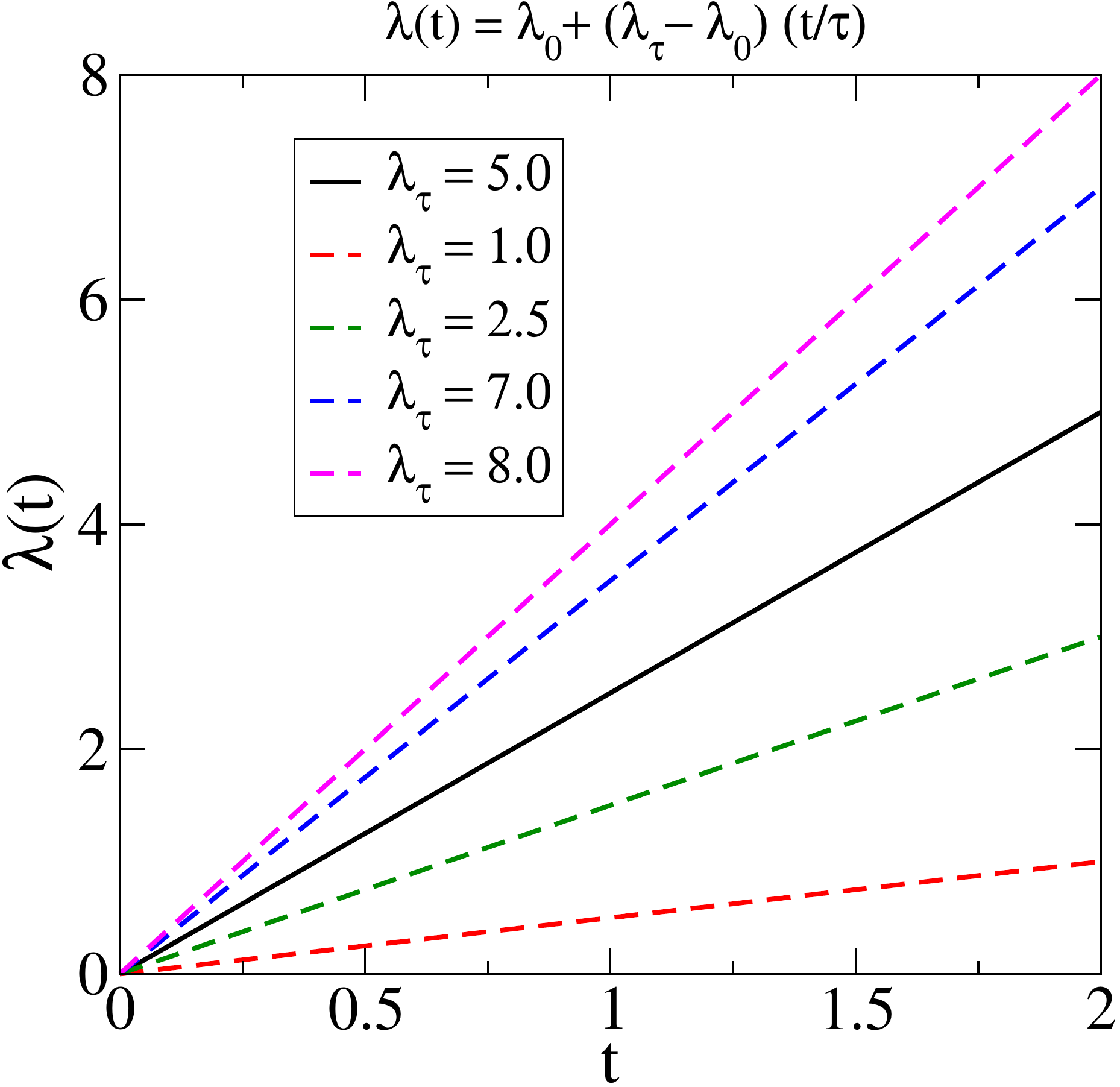}
    \caption{}
  \end{subfigure}
  \hspace{0.5cm}
  \begin{subfigure}{0.4\linewidth}
    \centering
    \includegraphics[width=7cm,height=7cm]{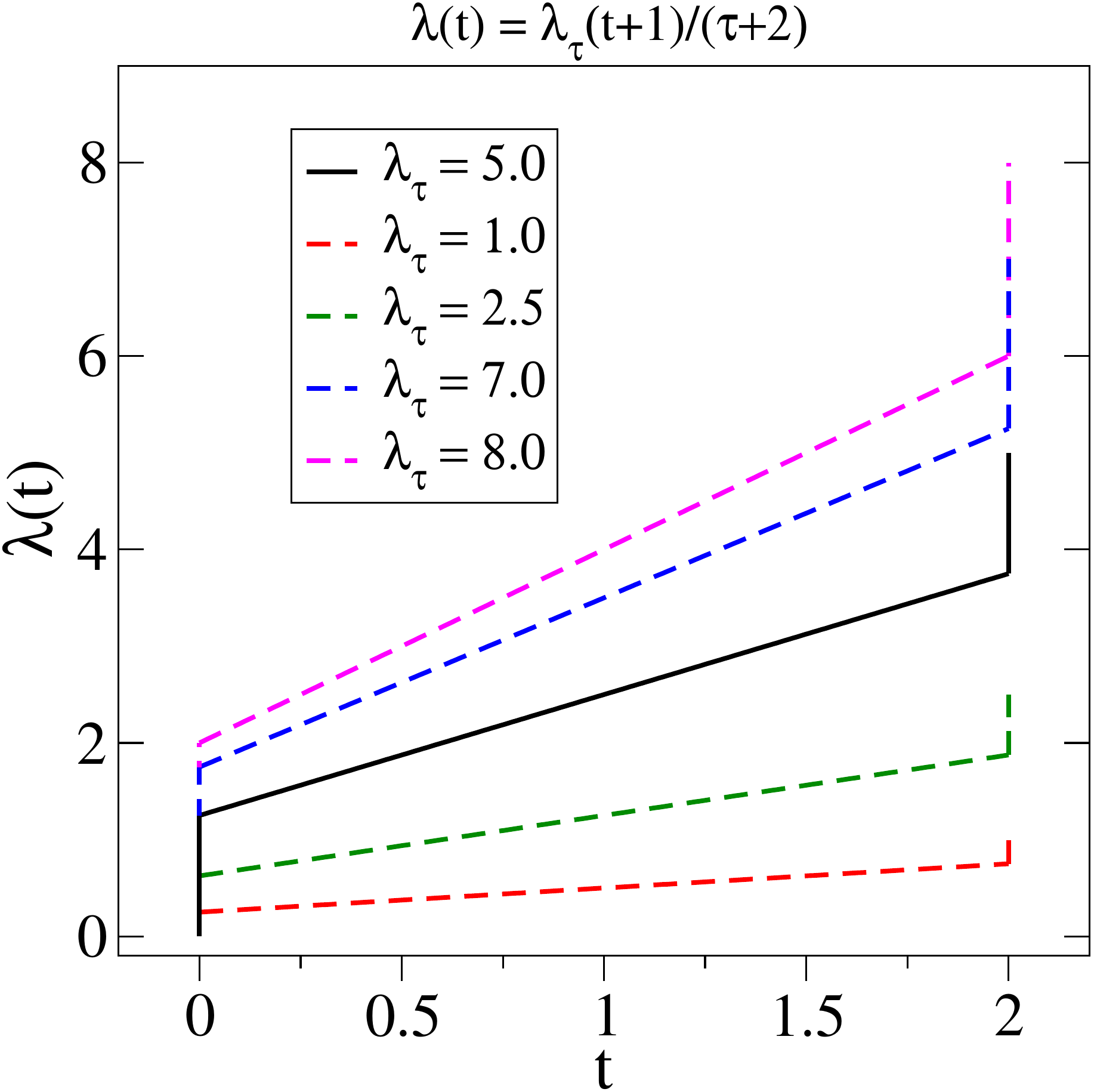}
    \caption{}
  \end{subfigure}
  \caption{ Example of (a) Protocol (1)  and (b) protocol (2). Unlike protocol (1) for protocol (2) there are sudden jumps at initial and final times, as discussed in the text. But $\Delta F=0$ for both the protocols. Graphs are plotted for  $\lambda_0=0$, $\tau=2.0$ for different $\lambda_\tau$.} 
  \label{prots}
\end{figure*}

In the case of optimal protocol, where the initial and final values of the parameter are given, there must be jumps at initial and final times as shown in \cite{sei07_prl}. This is not required in the ramp protocol, where the initial and final values of the protocol are connected simply by a straight line (obviously, this is \emph{not} an optimal protocol, since we are not allowing for the jumps to occur). However, in either case, as long as protocol is a simple dragging of the center of the trap, the free energy remains zero and the functional form of the protocol is immaterial. Fig. (\ref{prots}) shows an example of such protocols. 

To study how the convergence of the Jarzynski equality is affected when the protocol is randomized, we write the  Jarzynski equality as
\begin{eqnarray}
 \lim_{N\to\infty} \frac{1}{N}\sum_{i=1}^N e^{-\beta W_i} &= e^{-\beta\Delta F}.
\end{eqnarray}
Here, the index $i$ labels the individual realizations. 
In practice,  the relation ``converges'' if $N$ is large enough, so that we can obtain an accurate value (up to the allowed tolerance level) of $\Delta F$, that agrees with the ``correct'' value that emerges in the $N\to\infty$ limit (which can be calculated analytically in some simple cases). We would use different random distributions to sample the parameters of our protocol and calculate the minimum value of $N=N_c$ at which the relation converges. The distribution corresponding to which $N_c$ is smaller, is practically more useful, since we can obtain results in quicker time.

We will further find the convergence of the mean work obtained from simulation with the theoretically predicted value of mean work. In the next section, we calculate the theoretical expressions for the mean works with which we wish to compare the numerically obtained values.

\subsection{Ramp protocol}

For the ramp form of protocol, we first note that the work done will be given by \cite{sek98}
\begin{eqnarray}
  W = \int_0^\tau \pd{H(x,t)}{\lambda}\dot\lambda dt = - k\int_0^\tau [x(t)-\lambda(t)]\dot\lambda(t) dt\nn\\
  \av W = - k\int_0^\tau [\av{x(t)}-\lambda(t)]\dot\lambda(t) dt,
  \label{def:W}
\end{eqnarray}
where the angular brackets $\av{\cdot}$, denote the average over thermal noise.

Let us choose the initial value of the parameter to be fixed: $\lambda_0=0$, and the form of the protocol to be $\lambda(t) = \lambda_\tau t/\tau$. The parameter $\lambda_\tau$ is sampled from a random distribution, which we choose to be either uniform, Gaussian or exponential:
\begin{eqnarray}
  P(\lambda_\tau) &=& \frac{1}{\sqrt{2\pi\sigma_{\lambda}^2}}~e^{-(\lambda_\tau - \bar{\lambda}_\tau)^2/2\sigma_{\lambda}^2}\hspace{0.5cm}~\mbox{(Gaussian)};\\
  &=& 1 \hspace{4.2cm}~\mbox{(Uniform)};\\
  &=& (1/\bar{\lambda}_{\tau})~ e^{-\lambda_{\tau}/\bar{\lambda}_{\tau}}
  \hspace{1.8cm}~\mbox{(Exponential)}.
  \label{protocols}
\end{eqnarray}
For all distributions we choose $\bar{\lambda}_{\tau}=5$ as mentioned earlier. For simplicity, we have chosen the width of the Gaussian distribution $\sigma_{\lambda}$, to be unity. Similarly, we have chosen the domain of the uniform distribution to be unity, so that the sampled random numbers are in $(a,a+1)$. The value of $a$ depends on the specific example at hand. In our case, $a=4.5$, so that the mean is 5, which is also the value of $\lambda_\tau$ for the fixed protocol. We also note that for the exponential distribution, mean is equal to the standard deviation. The mean position is given by (here $\av{\cdot}$ represents averaging over the thermal noise, whereas the overbar represents average over $P(\lambda_\tau)$.)
\begin{eqnarray}
  \overline{\av{x(t)}} &= \frac{\overline{\lambda_\tau}}{\tau}(e^{-t}+t-1).
\end{eqnarray}
The mean work is given by (see Eq.\ref{def:W})
\begin{eqnarray}
  \overline{\av{W(\tau)}} &= \left(\frac{\overline{\lambda_\tau^2}}{\tau^2}\right)(e^{-\tau}+\tau-1).
  \label{Wmean_ramp}
\end{eqnarray}

As mentioned earlier the protocol does not smoothly extend to the given initial and final values of the protocol (which are fixed), so there are sudden jumps at the ends in order to abide by this constraint. This introduces an added contribution to the work. However, in the above case (dragging the center of trap), the value of equilibrium change in free energy identically vanishes, irrespective of the initial and final values of protocol. So we do not adhere to a fixed value of $\lambda_\tau$ in our analysis.

\subsection{Optimal protocol}

If we use the protocol used in \cite{sei07_prl}, then it is necessary to fix the final value of the parameter, a principle on which the derivation of the form of optimal protocol is based. In this case, the work done is given by 
\begin{eqnarray}
  \overline{\av{W(\tau)}} &= \frac{\overline{\lambda_\tau^2}~ \tau}{(\tau+2)^2} + \overline{\av{W_{end}}},
\end{eqnarray}
where $\overline{\av{W_{end}}}$ is the contribution coming due to the jumps at the initial and final points of the protocol. Below we calculate these contributions.
At the jumps, the work done on the system is given by the change in internal energy (since no heat is dissipated in a sudden quench). At the initial time $t=0$, we then have
\begin{align}
  W_{ini} &= \frac{1}{2}k[x_0-\lambda(0^+)]^2 - \frac{1}{2} k[x_0-\lambda_0]^2\nn\\
          &= \frac{1}{2} k (\lambda^2(0^+) - \lambda_0^2)~-~kx_0\lambda(0^+)~+~k x_0\lambda_0,
\end{align}
Using the fact that $\lambda_0=0$, averaging over the noise, and using $\av{x_0}=0$, the right hand side becomes simply $k\lambda^2(0^+)/2$. Now, integrating over the protocol distribution, we get
\begin{align}
  \overline{\av{W_{ini}}} &= \frac{k\overline{\lambda_\tau^2}}{2(\tau+2)^2}.
\end{align}
Similarly, at the final time $t=\tau$,
\begin{align}
  W_{fin} &= \frac{1}{2}k[x_\tau-\lambda_\tau]^2 - \frac{1}{2}k[x_\tau-\lambda(\tau^-)]^2\nn\\
          &= \frac{1}{2}k\lambda_\tau^2 ~\frac{2\tau+3}{(\tau+2)^2}  -kx_\tau~\frac{\lambda_\tau}{\tau+2},
\end{align}
where we have used $\lambda(t)=\left(\frac{t+1}{\tau+2}\right)\lambda_\tau$ (see \cite{sei07_prl}). In \cite{sei07_prl}, it was also shown that the position of the particle, when averaged over thermal noise, is given by $\av{x(t)} = \lambda_\tau t/(\tau+2)$. Therefore, after averaging over thermal noise, the right hand side becomes $\frac{3k\lambda_\tau^2}{2(\tau+2)^2}$. Thus
\begin{align}
  \overline{\av{W_{fin}}} &= \frac{3k\overline{\lambda_\tau^2}}{2(\tau+2)^2}.
\end{align}
Finally, adding both the contributions we get
\begin{align}
  \overline{\av{W_{end}}} &= \overline{\av{W_{ini}}} + \overline{\av{W_{fin}}} = \frac{2k\overline{\lambda_\tau^2}}{(\tau+2)^2}.
\end{align}
Setting $k=1$ in accordance with our choice, the total contribution of jumps, when averaged over noise as well as $P(\lambda_\tau)$ is given by 
\begin{eqnarray}
  \overline{\av{W_{end}}} &= \frac{2~\overline{\lambda_\tau^2}}{(\tau+2)^2}\ ,
                   \label{Wjumps}
\end{eqnarray}
thus
\begin{eqnarray}
\overline{\av{W(\tau)}} &= \frac{\overline{\lambda_\tau^2}}{(\tau+2)}\ .
\label{Wmean_opt}
\end{eqnarray}

We also note that for both the cases discussed above the work distributions are Gaussian for fixed protocols, and they satisfy Eqs. (\ref{JE}) and (\ref{CFT}). Then it is easy to show that the variance $\sigma_W^2 = 2k_BT \langle W\rangle $. However in case of protocols chosen from a
distribution $P(\lambda)$, work distributions are non-Gaussian and $P(W)$ can be calculated using
the Bayes' theorem:
\[P(W)= \int d\lambda ~P(\lambda)~ P(W|\lambda).\]
This in general can not be integrated to get a closed form. But $P(W)$ still satisfies Eqs. (\ref{JE}) and (\ref{CFT}).

\section{Numerical simulations}

\subsection{Uncorrelated protocol noise}

We first study the case of dragging protocol, where the equation of motion is given by Eq. (\ref{lang_drag}). 
We have used the velocity Verlet algorithm by Ermak \cite{ermak, allen_tildeslay} to integrate our equations of motion with time step $dt\sim 10^{-4}$, and have generated $\sim 10^5$ trajectories to construct a full ensemble.

Using the Jarzynski equality, $\av{e^{-\beta W}} = e^{-\beta\Delta F}$, one can calculate $\Delta F$ for each case. Ideally, an infinitely large number of trajectories need to be considered for taking the average appearing on the left hand side, but practically for a large enough number of trajectories we can obtain a convergence that is consistent with our required tolerance. Thus, if for a given protocol, if we find the value $\Delta F$ so obtained as a function of the ensemble size $N$, then an increase in $N$ would lead to a more accurate result.

If the analytical value of $\Delta F$ is known (which, for the dragging protocols, are zero), then we can find the required number $N=N_c$ at which the numerically obtained value falls within the tolerance range. A protocol that requires very high $N_c$ is much less efficient (in terms of convergence) when compared to one that provides convergence much faster. We set $\lambda_0=0$ and $\bar\lambda_\tau=5$ throughout. Similarly, one can compute the number of trajectories required for the convergence of mean work, whose expressions are given by equations Eq. (\ref{Wmean_ramp}) and Eq. (\ref{Wmean_opt}).

\begin{figure*}[!htbp]
    \begin{subfigure}{0.4\linewidth}
  \centering
  \includegraphics[width=7.5cm]{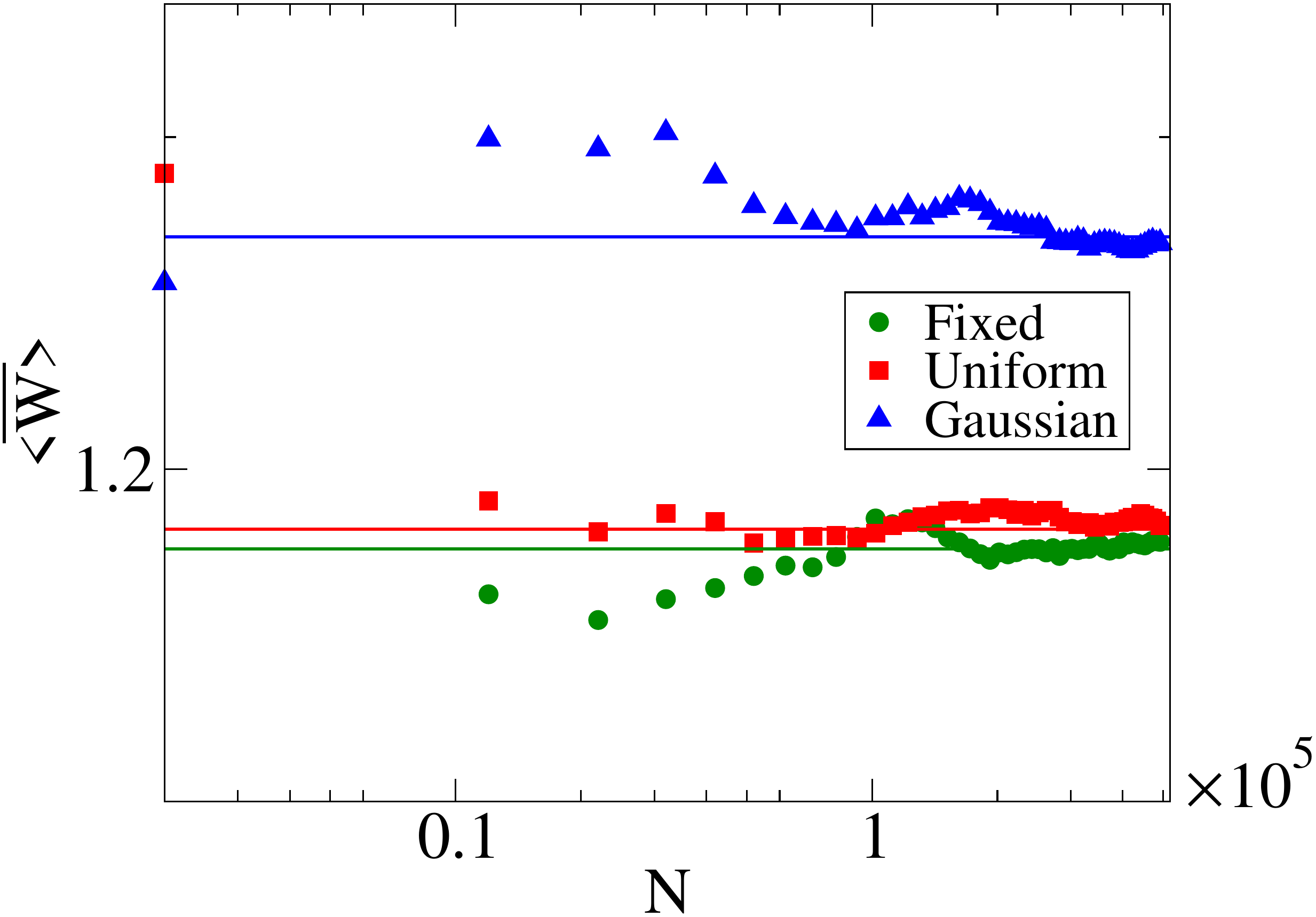}
    \caption{}
    \end{subfigure}
 \hspace{1.0cm}
 \begin{subfigure}{0.4\linewidth}
   \centering
   \includegraphics[width=8cm,height=5.5cm]{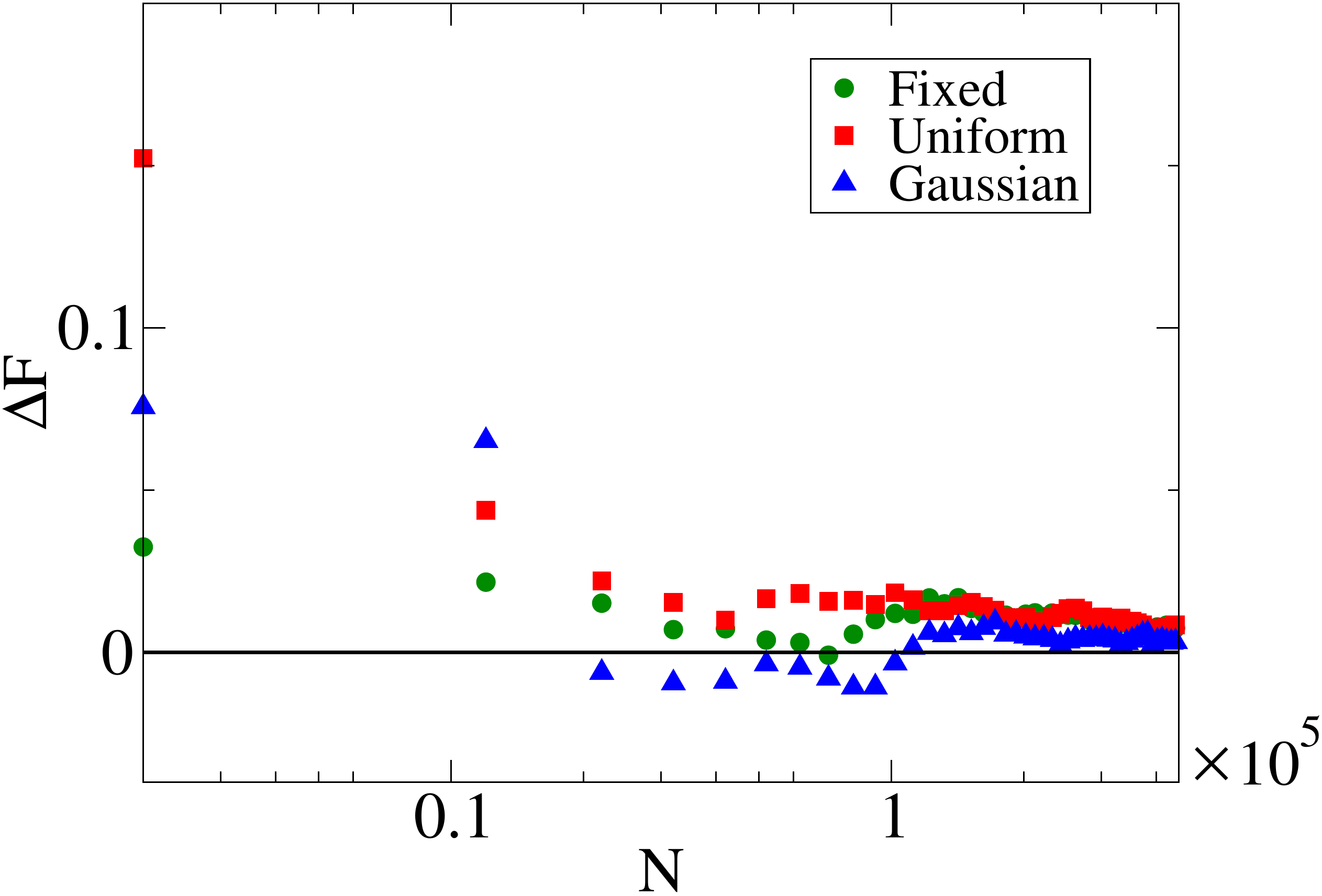}
   \caption{}
 \end{subfigure}
 \caption{(a) $\overline{\av W}$ vs $N$ and (b) $\Delta F$ vs $N$ for protocol (1) when the protocol is (i) fixed (green filled circles), (ii) uniform (red filled squares), (iii) Gaussian (blue filled triangles). In the left panel solid lines correspond to analytical values calculated from Eq. $(\ref{Wmean_ramp})$. The parameters used are: $k=1, ~\gamma=1, ~T=1$, $\tau=20$, $\bar\lambda_\tau=\lambda_\tau^{fix}=5$. Same parameters have been used in all the subsequent plots. The $N$-axis values are scaled down by a factor of $10^5$, as has been denoted in the plot by the multiplicative factor beside the axis.}
 \label{fig:W_N_ramp}
\end{figure*}

\begin{figure*}[!htbp]
    \begin{subfigure}{0.4\linewidth}
  \centering
  \includegraphics[width=7.5cm]{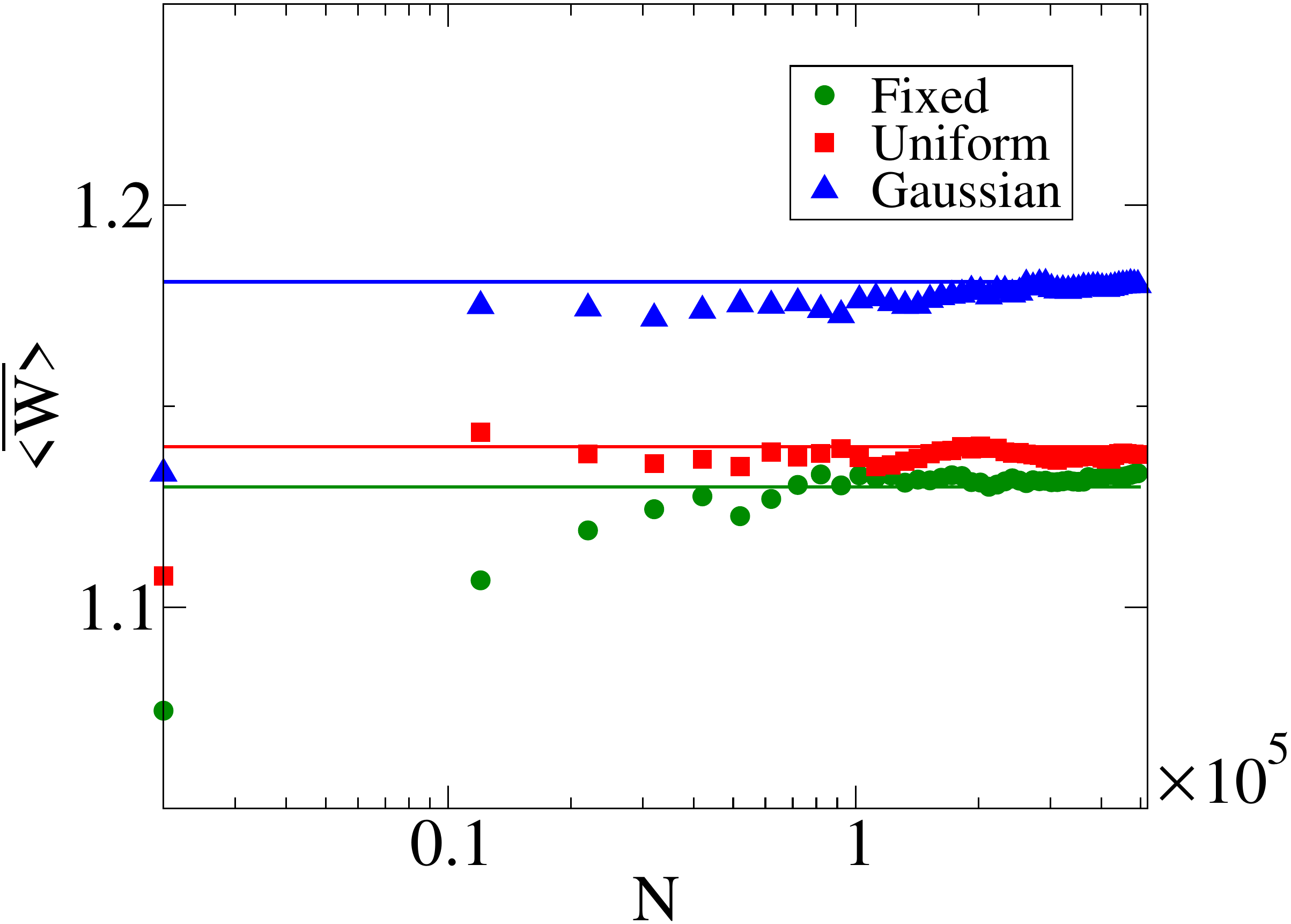}
    \caption{}
    \end{subfigure}
 \hspace{1.0cm}
 \begin{subfigure}{0.4\linewidth}
   \centering
   \includegraphics[width=8cm,height=5.2cm]{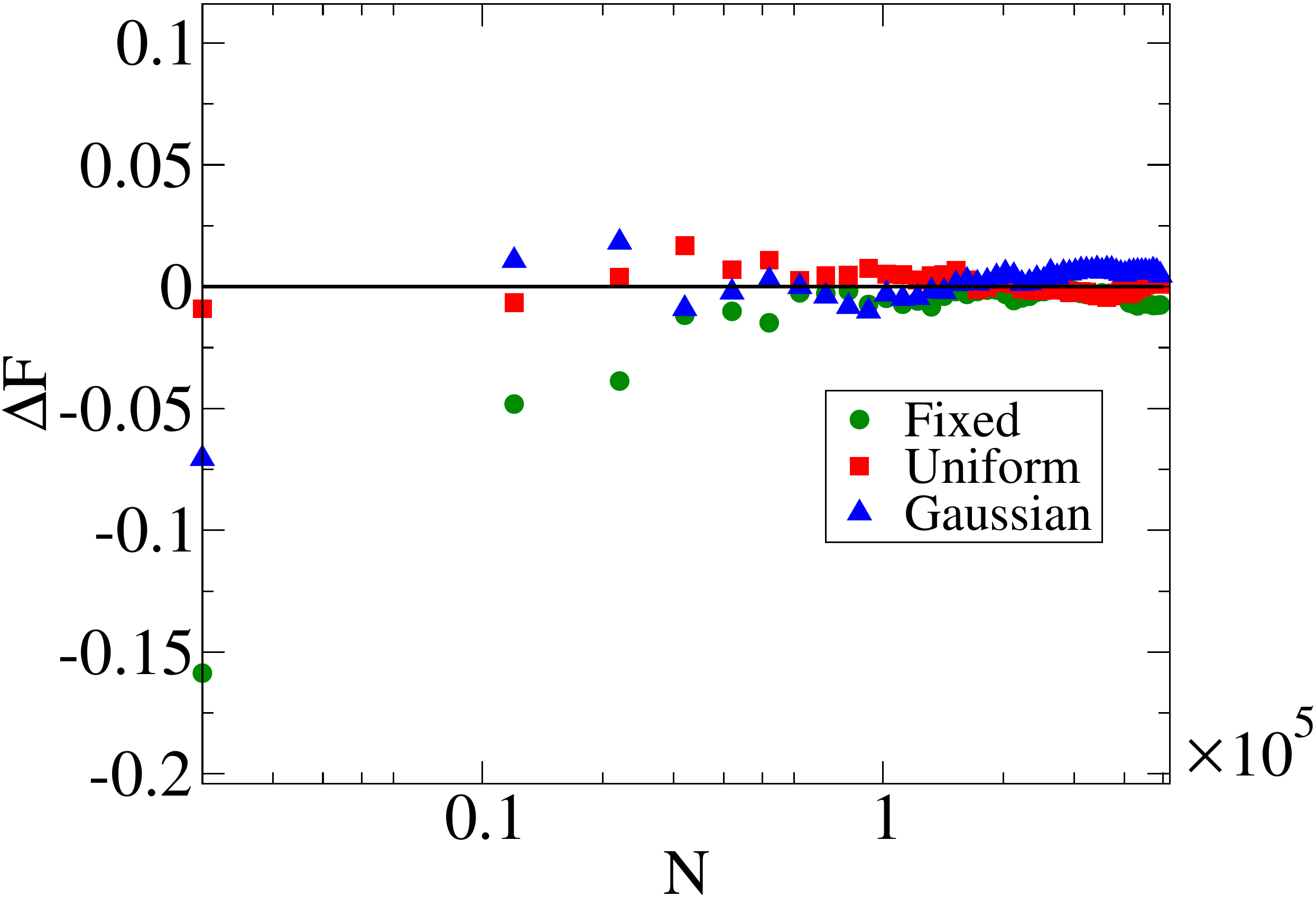}
   \caption{}
 \end{subfigure}
 \caption{(a) $\overline{\av W}$ vs $N$ and (b) $\Delta F$ vs $N$ for protocol (2) when the protocol is (i) fixed (green filled circles), (ii) uniform (red filled squares), (iii) Gaussian (blue filled triangles). In the left panel solid lines correspond to analytical values calculated from Eq. $(\ref{Wmean_opt})$. Parameters are same as in figure \ref{fig:W_N_ramp}. The $N$-axis values are scaled down by a factor of $10^5$, as has been denoted in the plot by the multiplicative factor beside the axis.}
 \label{fig:W_N_opt}
\end{figure*}

In figure \ref{fig:W_N_ramp}, we show this behavior when the protocol is of type (1), but the parameter $\lambda_\tau$ is either fixed, or sampled from a uniform and a Gaussian distribution. When it is random, we set the parameters of the distribution such that $\bar\lambda_\tau=\lambda_\tau^{fix}=5$. In the case of uniformly distributed $\lambda_\tau$, we adjust the range of the distribution to $\lambda_\tau\in (4.5,5.5)$, so that the mean value is $\bar\lambda_\tau=5$. For the normally distributed $\lambda_\tau$, we simply set the peak of the distribution at 5. The theoretical mean values of work are given by $\overline{\av W}= 1.188$ (fixed protocol), 1.191 (uniform distribution) and 1.235 (Gaussian distribution).
\begin{figure*}[t]
    \begin{subfigure}{0.4\linewidth}
  \centering
  \includegraphics[width=7cm]{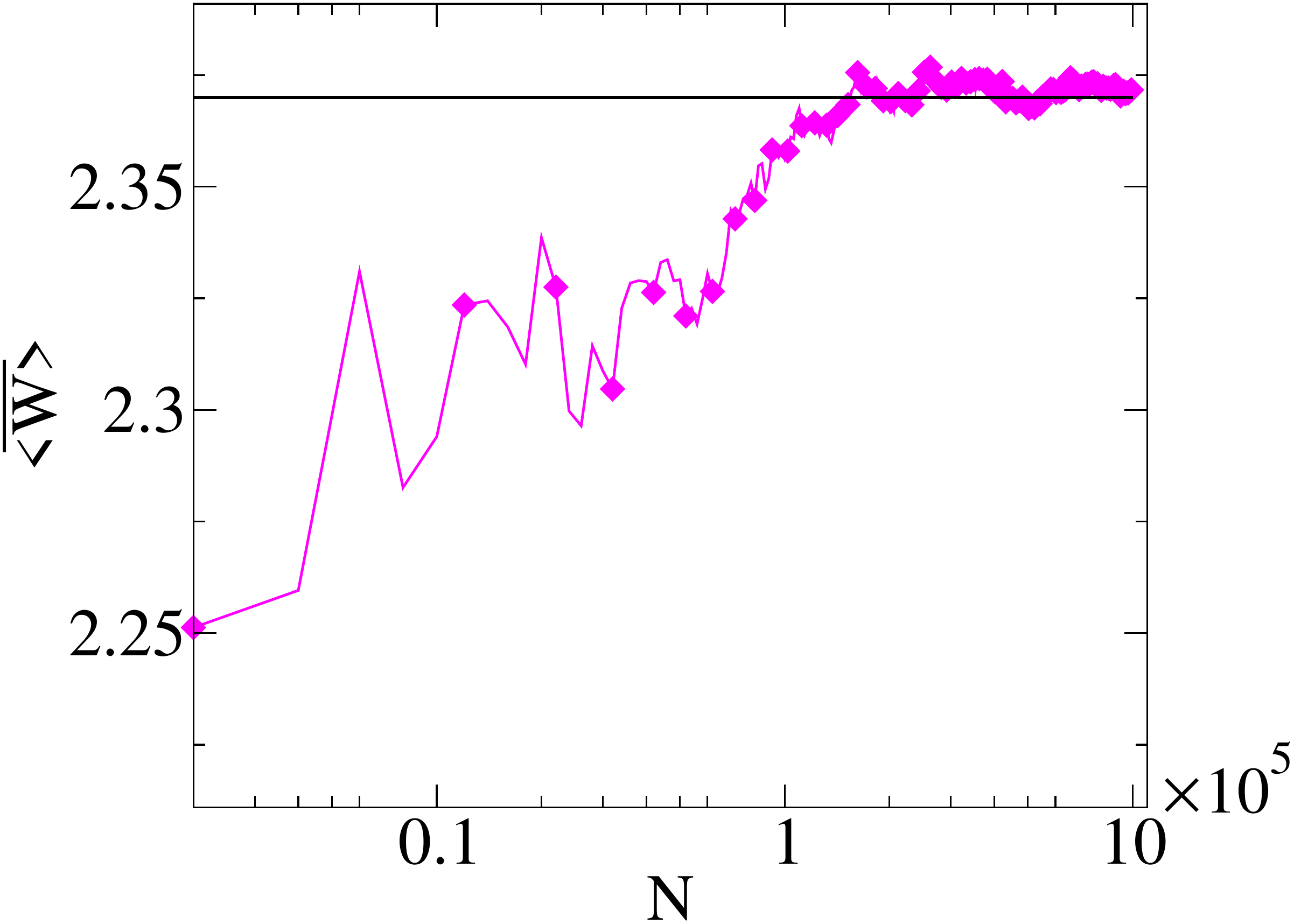}
    \caption{}
    \end{subfigure}
 \hspace{0.5cm}
 \begin{subfigure}{0.4\linewidth}
   \centering
   \includegraphics[width=7.8cm]{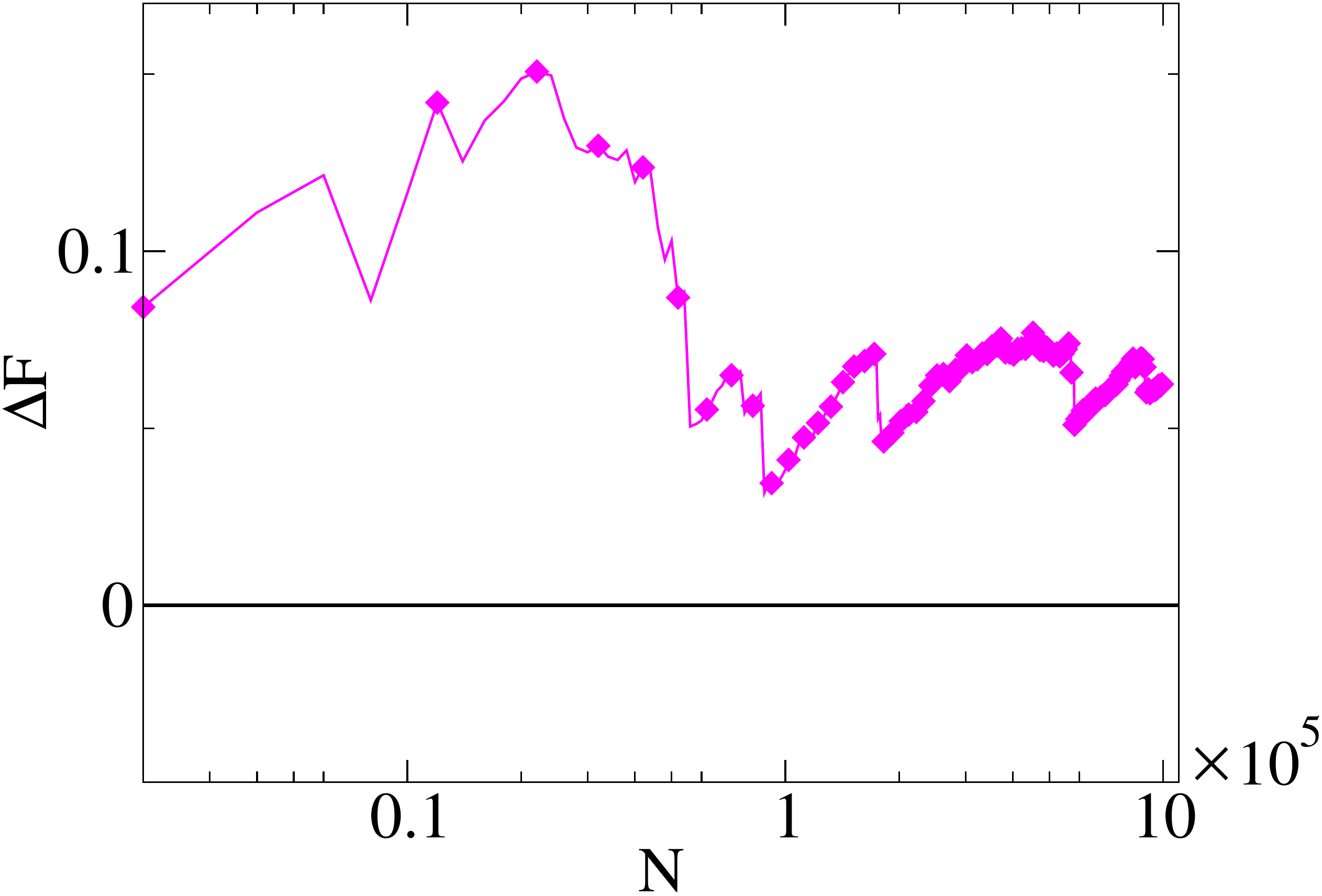}
   \caption{}
 \end{subfigure}
 \caption{(a) $\overline{\av W}$ vs $N$ and (b) $\Delta F$ vs $N$ for protocol $(1)$ when the parameter $\lambda_{\tau}$ chosen from an exponential distribution (see Eq.(\ref{protocols})). One can see that the exponential distribution does poorly in terms of convergence as compared to other distributions. Parameters are same as in figure \ref{fig:W_N_ramp}. The $N$-axis values are scaled down by a factor of $10^5$, as has been denoted in the plot by the multiplicative factor beside the axis.}
 \label{fig:exp_ramp}
\end{figure*}

\begin{figure*}[!htbp]
    \begin{subfigure}{0.4\linewidth}
  \centering
  \includegraphics[width=7.5cm,height=6cm]{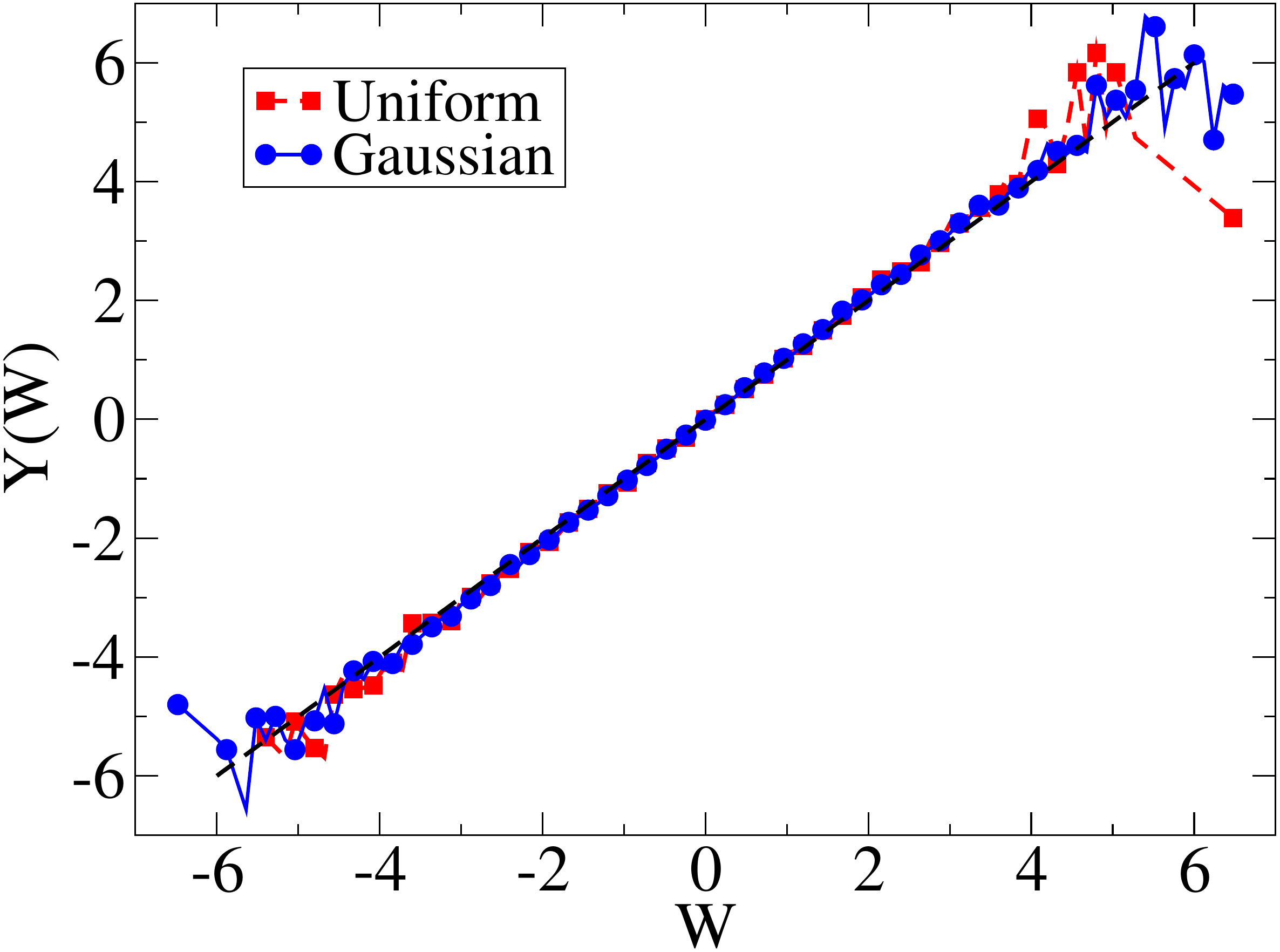}
    \caption{}
    \end{subfigure}
 \hspace{1.0cm}
 \begin{subfigure}{0.4\linewidth}
   \centering
   \includegraphics[width=7.5cm,height=6cm]{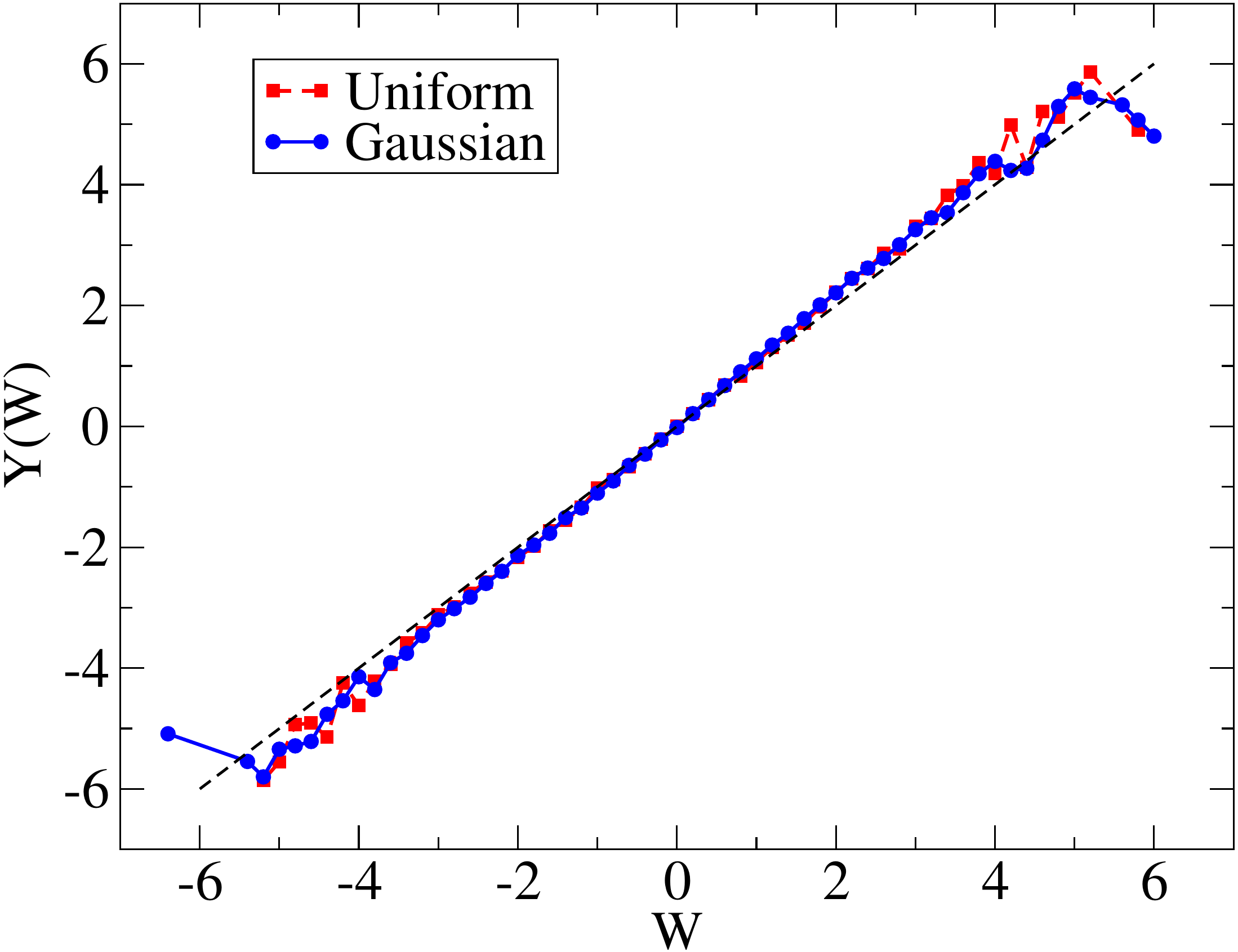}
     \caption{}
 \end{subfigure}
 \caption{(a) Symmetry functions for  protocol (1) with uniform and Gaussian distributions of $\lambda_\tau$. (b) Corresponding symmetry functions for protocol (2). The ensemble used for the purpose contained $N=10^5$ realizations, other parameters being same as in figure \ref{fig:W_N_ramp}.}
 \label{fig:symfunc}
\end{figure*}

In figure \ref{fig:W_N_opt}(a), where the convergence of mean work has been tested for protocol (2) i.e. the optimal protocol, the scenario for the convergence to theoretical value remains the same as in the case of ramp protocol, i.e. the values of $N_c$ are comparable in all the three cases and are roughly $N\approx 1\times 10^5$. The theoretical values of $\overline{\av W}$ are 1.136 (fixed protocol), 1.140 (uniform distribution) and 1.181 (Gaussian distribution).
  However, as we observe from figure \ref{fig:W_N_opt}(b) for $\Delta F$, we find that the convergences for uniform and Gaussian distributions become appreciable only beyond $N\approx 10^5$. We will observe later (see figure \ref{fig:symfunc}) that plotting the symmetry function is a better way to find $\Delta F$ efficiently.

\begin{figure*}[!htbp]
    \begin{subfigure}{0.4\linewidth}
  \centering
  \includegraphics[width=7cm]{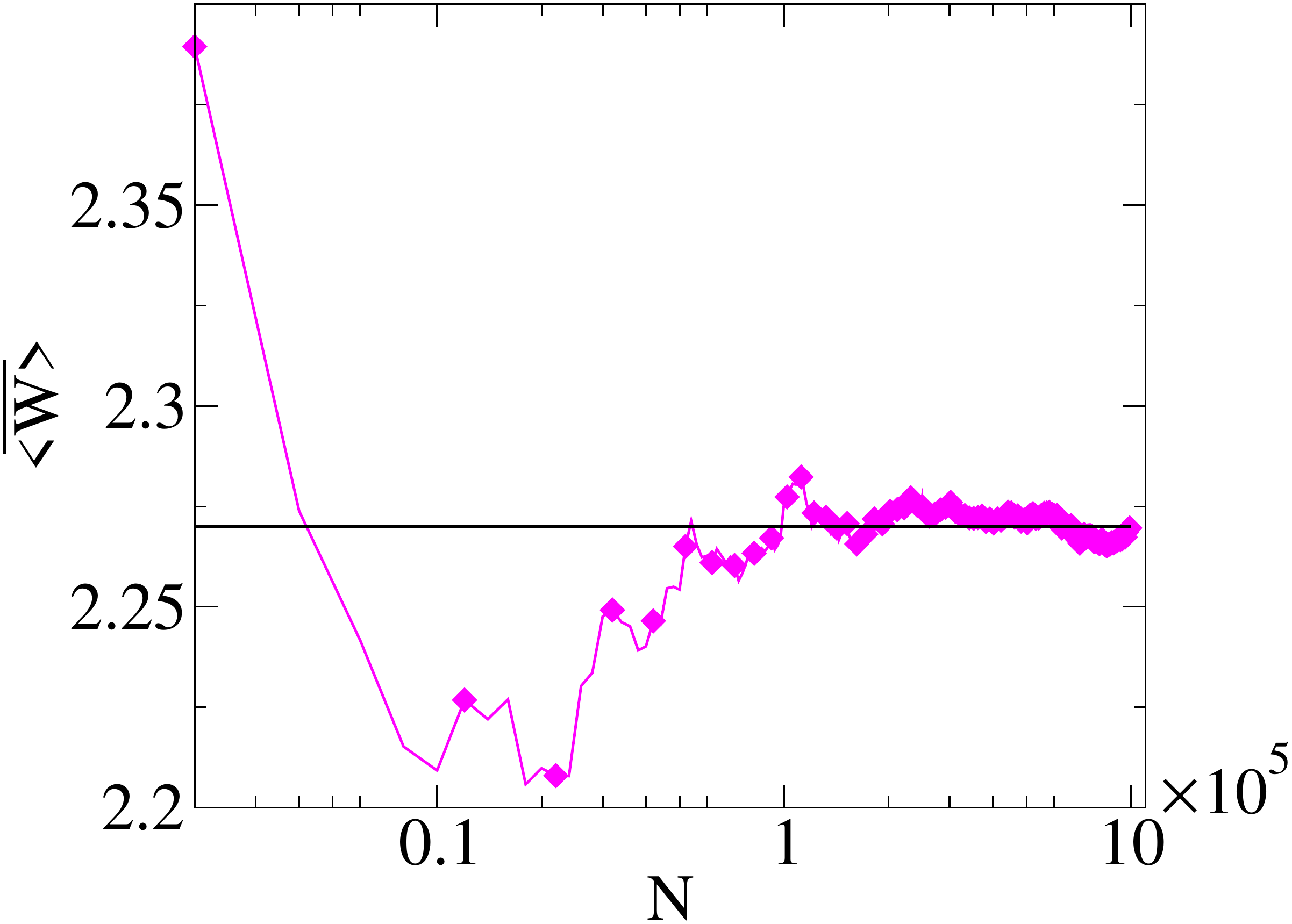}
    \caption{}
    \end{subfigure}
 \hspace{0.5cm}
 \begin{subfigure}{0.4\linewidth}
   \centering
   \includegraphics[width=7.7cm]{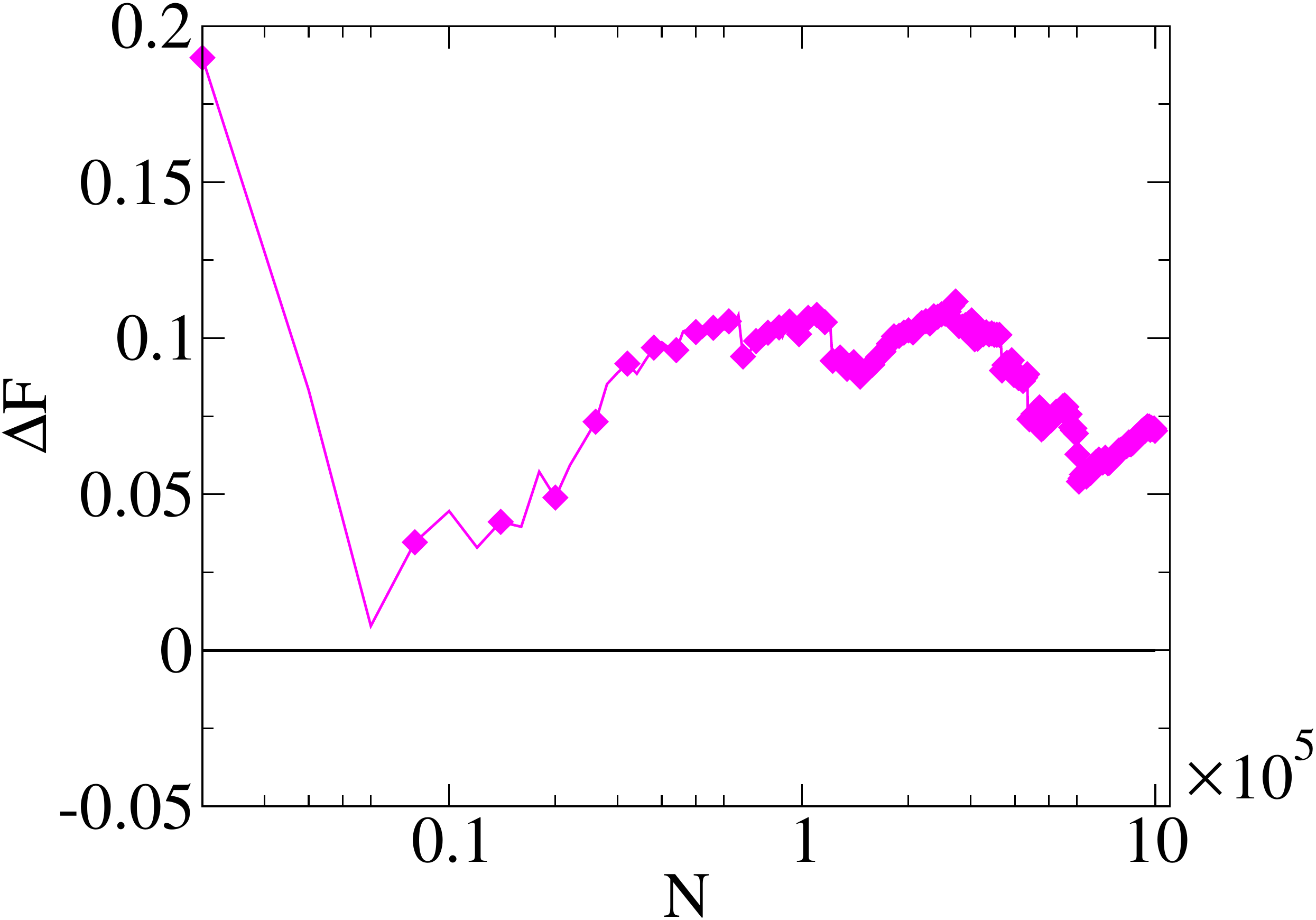}
   \caption{}
 \end{subfigure}
 \caption{(a) $\overline{\av W}$ vs $N$ and (b) $\Delta F$ vs $N$ for protocol $(2)$ when the parameter $\lambda_{\tau}$ chosen from an exponential distribution (see Eq.(\ref{protocols})). One can see that the exponential distribution does poorly in terms of convergence as compared to other distributions, even in case of optimal protocol. Parameters are same as in figure \ref{fig:W_N_ramp}. The $N$-axis values are scaled down by a factor of $10^5$, as has been denoted in the plot by the multiplicative factor beside the axis.}
 \label{fig:exp_opt}
\end{figure*}

Figure \ref{fig:exp_ramp}(a) and (b) show the convergence of mean work and free energy for protocol (1), respectively, when $P(\lambda_\tau)$ is exponentially distributed. Again, the convergence of mean work is appreciable roughly beyond $N\approx 2\times 10^5$. The convergence of free energy change is poor, even for $N\approx 10^6$, an obvious consequence of the large variance ($=5$) as compared to uniform and Gaussian distributions. The solid line in fig. \ref{fig:exp_ramp}(a) corresponds to the analytical value $\overline{\av W} = 2.370$. Figure \ref{fig:exp_opt} (a) and (b) displays the similar plots for protocol (2), for exponentially distributed $\lambda_\tau$. This time, the convergence in mean work appears to be quicker (figure \ref{fig:exp_opt}(a)), although that in the free energy change (figure \ref{fig:exp_opt}(b)) is again quite inefficient. The solid line in figure \ref{fig:exp_opt}(a) corresponds to the analytical value $\overline{\av W} = 2.273$. 

Finally, we turn our attention to the validity of Crooks Fluctuation Theorem for work (see Eq. (\ref{CFT})). This can be done by taking the logarithm of Eq. (\ref{CFT}):
\begin{eqnarray}
  Y(W) \equiv k_BT\ln\frac{P(W)}{\tilde P(-W)} = W-\Delta F.
\end{eqnarray}
The quantity on the left hand side is called the \emph{symmetry function}, and according to the above relation, if CFT is valid, must yield a straight line of slope unity as a function of $W$.

In figure \ref{fig:symfunc}, we have plotted the symmetry functions for uniform distribution and Gaussian distributions $P(\lambda_\tau)$ corresponding to each type of protocol: ramp (protocol (1), figure \ref{fig:symfunc}(a)) and optimal (protocol (2), figure \ref{fig:symfunc}(b)). In (a), we again find excellent agreement for uniform protocol. Even for the Gaussian protocol, we find very good agreement. In (b), we again find very good agreement for the protocol (2) having uniformly distributed $\lambda_\tau$. The Gaussian, however, shows deviations beyond $W=3$, which can be accounted for by the poor convergence of protocol (2) for large values of work (sampling of rare trajectories becomes difficult). Note that the convergence is not a problem for uniform protocol, since the values of $\lambda_{\tau}$ are bounded.
                    
\section{Numerical results when work distributions are non-Gaussian even for deterministic protocol}
\subsection{Numerical results for time-dependent trap strength}
In all the cases discussed above the work distributions are Gaussian. Here we give numerical results for a potential of the form $V_0(x)=\lambda(t) x^2/2$, where $\lambda(t)=\lambda_0 + A \sin(\pi t/\tau)$, is a half sinusoidal cycle. The equation of motion then becomes
\begin{equation}
  \label{sine}
  \gamma\dot x = -\lambda(t)~x~+~\xi(t).
\end{equation}

\begin{figure}[!htbp]
  \centering
  \includegraphics[width=9cm]{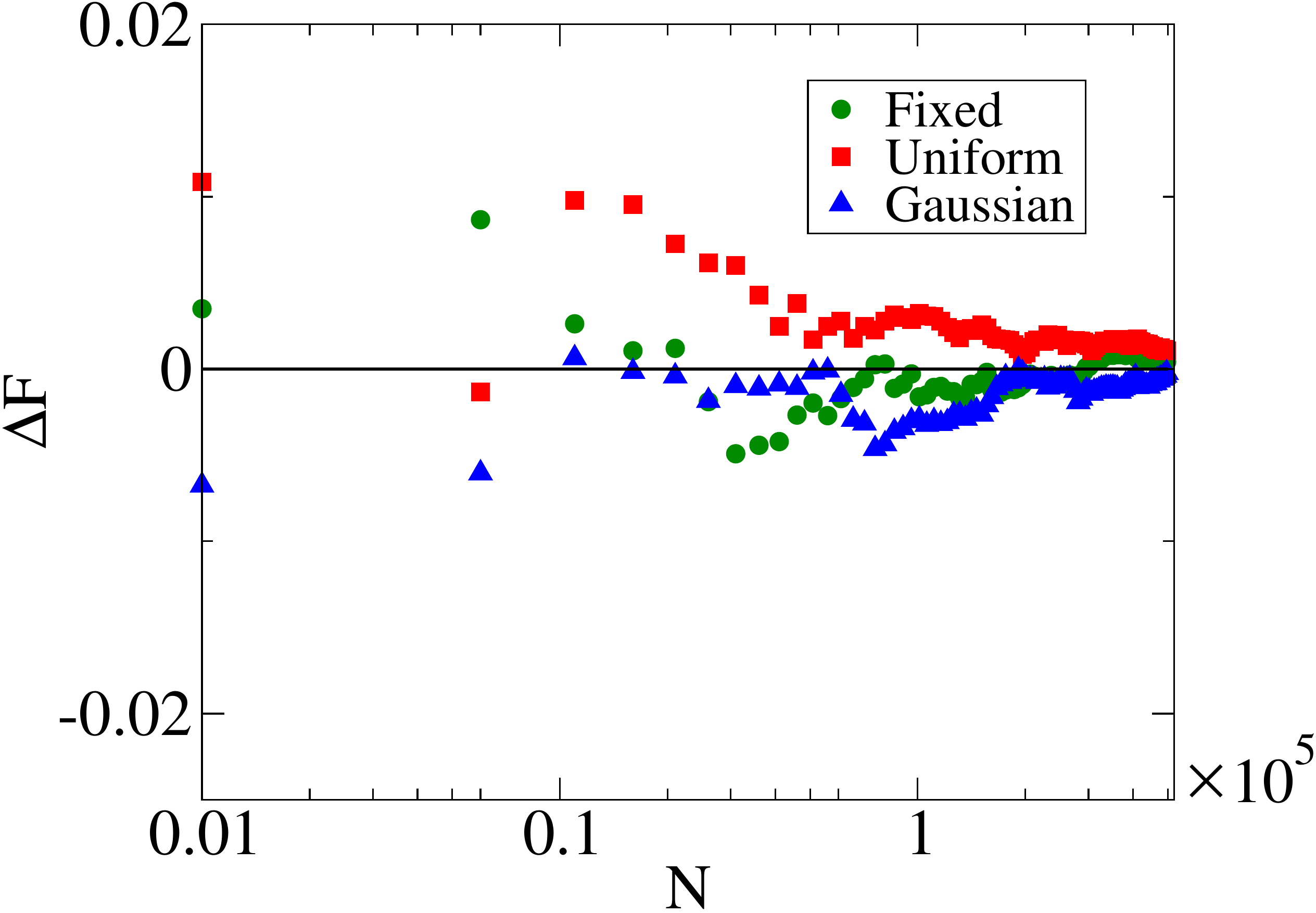}
  \caption{Plot of change in free energy, as a function of the number of realizations. The protocol is $\lambda(t)=\lambda_0 + |A| \sin(\pi t/\tau)$. The green filled circles are for fixed protocol, red filled squares for uniformly distributed $A$, while blue filled triangles are for normally distributed (with variance $5$) $\lambda_{\tau}$. Here, we choose $dt\sim 10^{-4}$, $T=1$, $\gamma=1$, $\tau=20$. For all distributions $\bar A=A_{fix}=5.0$. The $N$-axis values are scaled down by a factor of $10^5$, as has been denoted in the plot by the multiplicative factor beside the axis.}
  \label{fig:sintrap}
\end{figure}
The parameter $A$ is chosen from a given distribution as earlier. Unlike earlier, $\lambda_0\neq 0$ in this case to make sure that $\lambda(t) > 0$ for the full duration of the cycle. We choose $\lambda_{0} =1.0$. For uniform distribution we choose $\bar A=5$.  For Gaussian distribution we choose both mean and variance to be $5$. Note that for Gaussian distribution we also need to take $\lambda(t) = \lambda_0+|A|\sin(\pi t/\tau)$ to make sure $\lambda(t) > 0$ for the full duration of the cycle. The work done is given by
\[  W = \frac{1}{2}\int_0^\tau \dot{\lambda}(t)x^2(t) dt,\]
which is clearly non-Gaussian, being quadratic in $x$, even for fixed $A$.

Figure \ref{fig:sintrap} shows the convergences for free energy change (which is again zero, since $\lambda(\tau)=\lambda_0$ identically). It is clear that even with such a non-Gaussian distribution of work, the convergences are nearly same as in the case of the corresponding deterministic protocol.

We perform a more stringent verification of our results in the next section, where we replace the harmonic trap by a bistable potential.
\begin{figure}[!htbp]
  \centering
  \includegraphics[width=9cm]{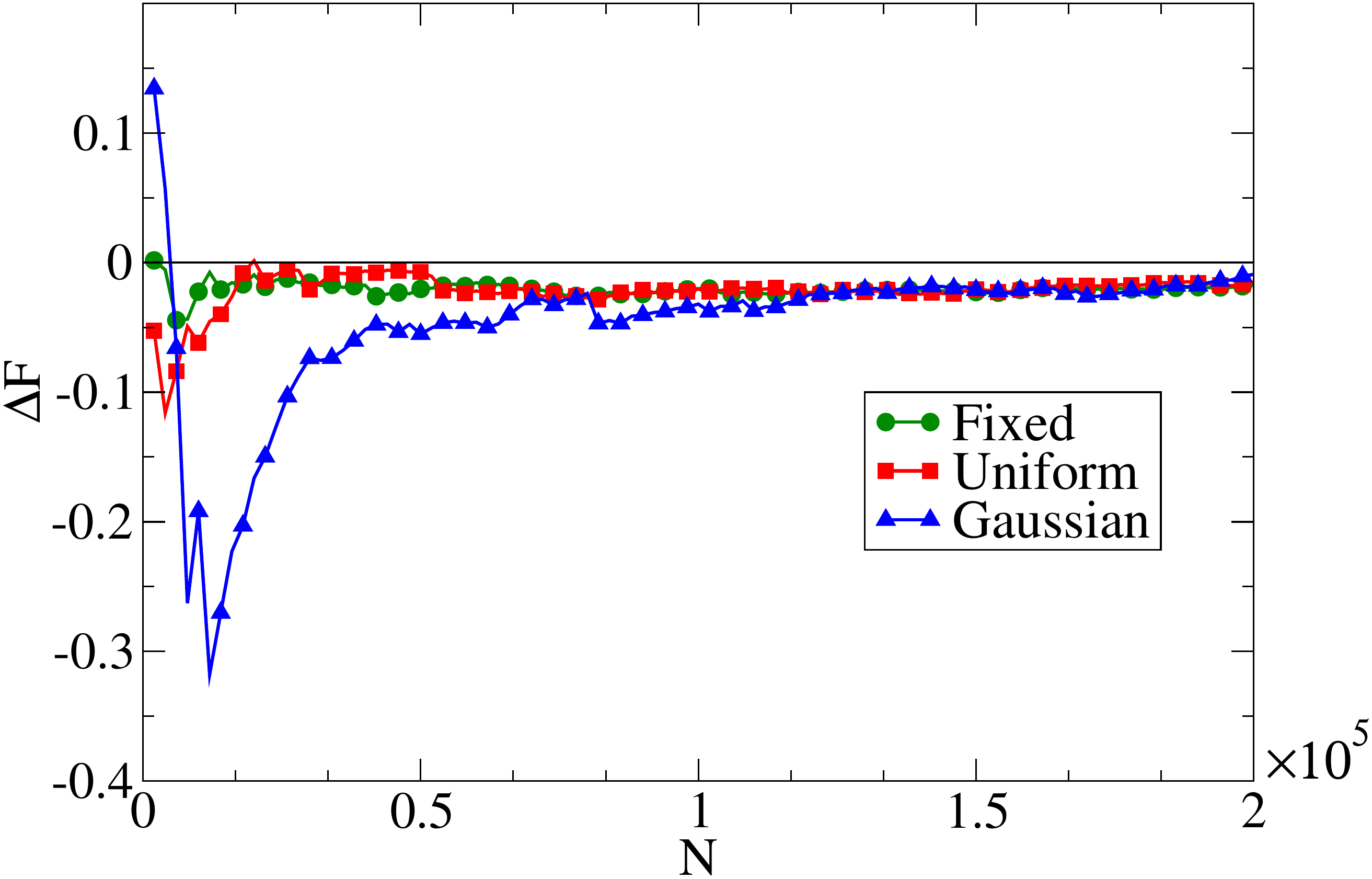}
  \caption{Plot of change in free energy, as a function of the number of realizations. The protocol is for $\bar A=A_{fix}=0.1$, $T=1$ and $\omega=1$. The green line with filled circles is for fixed protocol, red line with filled squares is for uniformly distributed $A$, while blue  line with filled triangles is for normally distributed (with variance unity) $A$. Here, we choose $dt\sim0.01$. The $N$-axis values are scaled down by a factor of $10^5$, as has been denoted in the plot by the multiplicative factor beside the axis.}
  \label{fig:doublewell}
\end{figure}

\subsection{Numerical results for bistable potential}
In order to check that the efficiency of random protocol is not a consequence of the simplistic form of quadratic potential, we now numerically simulate a bistable or double-well potential perturbed by a sinusoidal drive, of the form $V(x)=-x^2/2+x^4/4-A\sin(\omega t) x$. The Langevin equation now is
\begin{equation}
  \label{double}
  \gamma\dot x = x-x^3 + A\sin(\omega t)+\xi(t).
\end{equation}
The expression for work becomes
\[  W = -A \omega \int_0^\tau \cos(\omega t)x(t)dt.\]
Here, again the work distribution is clearly non-Gaussian in nature since the distribution of $x$ itself is non-Gaussian, even for fixed $A$.

The perturbation is carried out for one full time-period of the drive: $\tau=\tau_\omega=2\pi/\omega$, so that $\Delta F=0$ in the process. The results are shown in figure \ref{fig:doublewell}. We clearly observe that the blue curve (where $A$ is distributed normally with variance of unity, with the mean value fixed at $\bar A=A_{fix}=0.1$, $A_{fix}$ being the value of the parameter for a deterministic protocol) takes an extremely long time to converge, and as such is highly inefficient. Nevertheless, the uniform distribution does almost as good a job as the fixed one.

\section{Heuristic explanation for the observed behavior}

If we consider the averaging over protocols and trajectories as a double integral, we might naively expect that the convergence would require about $N_c=N_1 N_2$ realizations of the experiment, where $N_1$ is the number of trajectories required for convergence with a fixed protocol and $N_2$ is the size of the ensemble of protocols.
Thus, if $N$ realizations are needed for convergence under a fixed protocol, $\approx N^2$ realizations should be required for random protocols. However, we find that this is never the case, and convergence appears much earlier. In our opinion, the following reason accounts for this apparently unusual result. In general, all values of $\lambda_\tau$ sampled from a particular distribution need not require equal orders of $N_1$. For instance, in the Gaussian distribution, the values of $\lambda_\tau$ that are closer to $\bar\lambda_\tau$ would represent realizations where the protocol is close to being a fixed one, so that the convergence would occur earlier. On the other hand, those appearing at the tails would require an $N_1$ that is higher by an order of magnitude or more. However, since the weightage of these ``rare values'' is much less as compared to the ones that have values of $\lambda_\tau$ comparatively closer to $\bar\lambda_\tau$, the value of $N_c$ must be reduced considerably. Note that this happens clearly for the uniform and the Gaussian case, where the variances are equal to 0.083 and 1 respectively, which are small compared to $\bar\lambda_\tau$. 
For the  exponential distribution, however, the most probable value of $\lambda_\tau$  is appreciably different from its mean. Thus, a major contribution comes from values of $\lambda_\tau$ that are far from $\bar\lambda_\tau$, and due to this reason it is difficult to obtain convergence within the size of the ensemble that we have been able to simulate.

\begin{figure}[!ht]
 \centering
 \includegraphics[width=8cm]{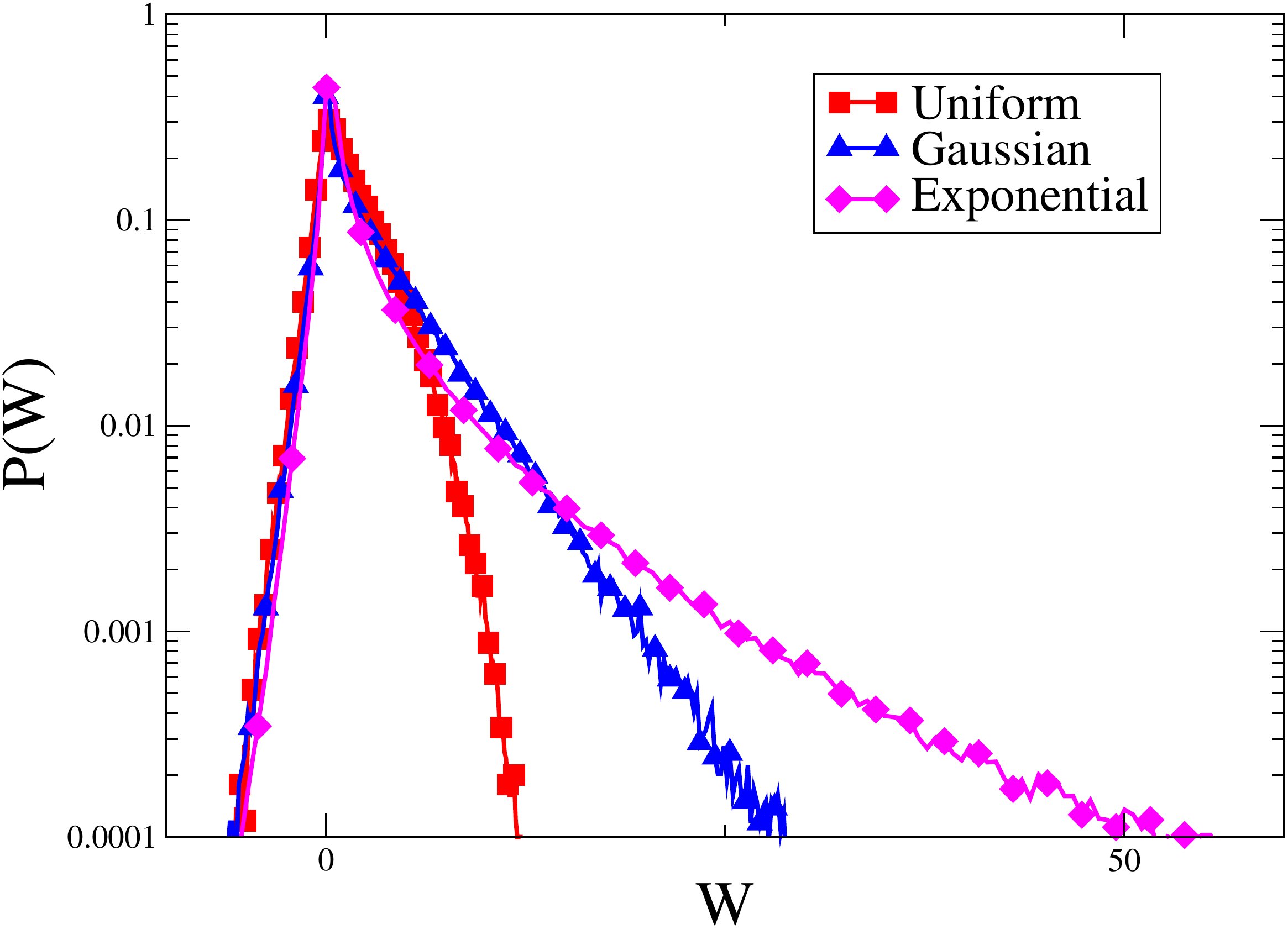}
 \caption{Plot of work distributions for the ramp protocol for three kinds of $P(\lambda_\tau)$, on a semi-log scale. The Gaussian protocol gives an exponential decay, while the uniform and the exponential ones show slower and faster decays in $P(W)$, respectively. The parameters are chosen such that the $\bar{\lambda}_{\tau}=5$
but $(1)$ Uniform: $\lambda_{\tau} \in [1,8]$, $(2)$ Gaussian: $\sigma_{\lambda} =5$ and $(3)$ Exponential: $\bar{\lambda}_{\tau} =5$ (as earlier). The parameters are chosen so that the deviation of $P(W)$ from Gaussianity and non exponential tails are clearly visible. Other parameters are same as Fig. (\ref{fig:W_N_ramp}).}
 \label{fig:Wdist}
\end{figure}

Our analysis corroborates this heuristic argument. We note that the approach could be useful, for example, when there is an inherent error in the control of the external parameter, so that to some extent the protocol gets randomized. The nature of work distributions have been shown in figure \ref{fig:Wdist}, where we find that the Gaussian randomness in protocol gives rise to an exponential distribution, the uniform one leads to faster decay and the exponential one leads to slower decay. This again shows that an exponential randomness in $\lambda_\tau$ requires a larger ensemble for convergence. In the Gaussian case, however, the rate of convergence will become slow if the variance is large as compared to the mean, which happens when the rate parameter $\alpha$ in $P(W)\sim e^{-\alpha |W|}$ (note that this can be different on either sides of the peak) becomes much smaller than unity \cite{sab11_epl}.

\section{Relation to incorrect work measurement}
We can also view the problem in the light of \cite{lah16_pre}, where instead of randomizing the protocol, there were random errors in the measurement of work itself.
In \cite{lah16_pre} it was shown that even if the work done in a process is measured inaccurately (might be due to low resolution of the apparatus measuring the particle's position), in several situations that occur frequently in experiments one can deduce the correct value of free energy change, using the symmetry functions obtained from the Crooks' work fluctuation theorem. We will show that the work done in our case can be expressed as the work done $W_{fix}$ using a fixed protocol, and an error term $\mathcal{E}$ (see eq. \eqref{final_connection} below). However, the error term turns out to be non-Gaussian in nature, which produces a more complex situation as compared to \cite{lah16_pre}, whose analytical techniques would be useful only if one of the terms in the error becomes negligible as compared to the other terms, as detailed below. 

Let us consider the ramp protocol ($\lambda(t)=\lambda_\tau t/\tau$), where $\lambda_\tau$ values are normally distributed. The work done is given by
\begin{eqnarray}
  W &= -k\int_0^\tau [x-\lambda(t)]\dot\lambda(t)dt = -k\int_0^\tau\left(x-\frac{\lambda_\tau t}{\tau}\right)\frac{\lambda_\tau}{\tau}dt.
\end{eqnarray}
Now, we can define the deviation of $\lambda_\tau$ from its mean as $\Delta\lambda_\tau$, so that $\lambda_\tau = \bar\lambda_\tau+\Delta\lambda_\tau$, so that
\begin{eqnarray}
  W &=&  -k\int_0^\tau\left[x-\frac{(\bar\lambda_\tau+\Delta\lambda_\tau) t}{\tau}\right]\left(\frac{\bar\lambda_\tau+\Delta\lambda_\tau}{\tau}\right)dt\nn\\
    &=& W_0 - k\int_0^\tau \left(x-\frac{\bar\lambda_\tau t}{\tau}\right)\frac{\Delta\lambda_\tau}{\tau}dt + \frac{k\bar\lambda_\tau}{\tau^2}\int_0^\tau \Delta\lambda_\tau tdt + \frac{k}{\tau^2}\int_0^\tau \Delta\lambda_\tau^2 tdt \nn\\
    &=& W_0+\mathcal{E}',
\end{eqnarray}
where $\mathcal{E}'$ gives the last three terms on the RHS. The above relation provides the general expression for work, \emph{irrespective} of the distribution of protocol. Here,
\begin{eqnarray}
  W_0 = -k\int_0^\tau \left(x-\frac{\bar\lambda_\tau t}{\tau}\right)\frac{\bar\lambda_\tau}{\tau}dt.
\end{eqnarray}
Note that $W_0$ is \emph{not} the work done by a fixed protocol, because of the dependence of $x$ on the protocol. Nevertheless, the solution for the Langevin equation Eq. (\ref{lang_drag}) is
\begin{eqnarray}
  x(t) &=& x_0e^{-kt/\gamma}+\frac{e^{-kt/\gamma}}{\gamma}\int_0^t \left[\frac{k(\bar\lambda_\tau+\Delta\lambda_\tau)}{\tau}t' + \eta(t')\right]e^{kt'/\gamma}dt'\nn\\
       &=& x_{fix}(t) ~+ \frac{ke^{-kt/\gamma}}{\gamma\tau}\int_0^t\Delta\lambda_\tau t'e^{kt'/\gamma}dt'\nn\\
       &=& x_{fix}(t) + \Delta x(t).
\end{eqnarray}
Here, $x_0$ is the value of the position of the particle at $t=0$, which follows a Boltzmann distribution. Therefore, $W_0$ can now be related to the work done for fixed protocol:
\begin{eqnarray}
  W_0 &= W_{fix} ~+~ \left( -k\int_0^\tau \left(\Delta x(t)-\frac{\bar\lambda_\tau t}{\tau}\right)\frac{\bar\lambda_\tau}{\tau}dt\right) \equiv W_{fix} + \Delta W.
\end{eqnarray}
Collecting all results, we have
\begin{eqnarray}
  W &= W_{fix} + \mathcal{E},
      \label{final_connection}
\end{eqnarray}
where $\mathcal{E} = \Delta W + \mathcal{E}'$.

However, if $\Delta\lambda_\tau$ is Gaussian (of zero mean in the present case), then we find that the last term in $\mathcal{E}'$ is non-Gaussian. As a result, a direct connection with \cite{lah16_pre} is not on offer. However, if the variance is small enough, this term can be ignored, and we should recover the results of \cite{lah16_pre}. 
\section{Conclusions}
The work fluctuation theorems use deterministic protocols that are exactly the same for all experimental realizations constituting the ensemble. In this work, we study how the results are affected when the protocol is not deterministic but has a random part, so that their time dependence is different for different experimental realizations. We find that if the initial and final values of the externally manipulated parameter are held fixed, then irrespective of the type of randomness incorporated, the Jarzynski Equality is satisfied when the ensemble is large enough. We find that even $N$ experiments (number of realization needed for convergence when a deterministic or fixed protocol is applied) suffice to provide a good convergence of our results, which is surprising.

To check the convergence of Jarzynski Equality, we have studied a dragged harmonic trap, given by Eq. \eqref{lang_drag}, where the equilibrium free energy is known to remain constant ($\Delta F = 0$). As a further check, we have analytically computed the mean works done in all the cases, and have observed the convergence of $\overline{\av W}$ obtained from numerics with the theoretical values, as a function of the ensemble size.

We have numerically tested our results in a couple of examples where the work distribution follows a non-Gaussian distribution. These cases include a harmonic trap of varying stiffness, and a bistable potential. In both cases, we find that the convergence rates with number of realizations are comparable.

We also believe that this study can be extended to steady state fluctuation theorems. This work is currently underway.  

\section{Acknowledgment}
SL thanks DST-SERB, India (sanction order no. ECR/2017/002607) for financial support. RM thanks C. Jarzynski for initial discussions on this topic. Authors also thank 
D. Lacoste for careful reading of the manuscript.

\end{document}